\documentclass[journal]{IEEEtran}

\usepackage{cite}
\usepackage[cmex10]{amsmath}
\usepackage{hyperref}
\usepackage{amssymb}
\usepackage{amsthm}
\usepackage{amssymb,amsthm,amscd}
\newcommand{\Z}{\mathbb{Z}}

\newtheorem{theorem}{Theorem}[section]
\newtheorem{lemma}[theorem]{Lemma}
\newtheorem{corollary}[theorem]{Corollary}
\newtheorem{proposition}[theorem]{Proposition}
\theoremstyle{definition}
\newtheorem{definition}{Definition}
\newtheorem{example}[theorem]{Example}
\newtheorem{remark}[theorem]{Remark}

\begin{document}
\title{\bf On $Z_p Z_{p^k}$-additive codes and their duality
\thanks{This is the peer reviewed version of the following article: M. She, R. Wu, D. S. Krotov, On $Z_p Z_{p^k}$-additive codes and their duality, IEEE Transactions on Information Theory, 
\url{https://doi.org/ 10.1109/TIT.2018.2883759}, pulshed online 30 November 2018.}
\thanks{This research is supported by National Natural Science Foundation of China (61672036), Excellent Youth Foundation of Natural Science Foundation of Anhui Province (1808085J20),
Technology Foundation for Selected Overseas Chinese Scholar, Ministry of Personnel of China (05015133) 
and the Program of fundamental scientific
researches of the Siberian Branch of the Russian Academy of Sciences No.I.1.1.
(No.0314-2016-0016).}
}
\author{%
Minjia Shi%
\thanks{%
     M. Shi is with the School of Mathematical Sciences,
     Anhui University, Hefei, Anhui, 230601, P. R. China
     (e-mail: smjwcl.good@163.com)%
}%
, Rongsheng Wu%
\thanks{%
     R. Wu is with the School of Mathematical Sciences,
     Anhui University, Anhui, 230601, P. R. China
     (e-mail: wrs2510@163.com)%
}%
, and Denis S. Krotov, 
\IEEEmembership{Member, IEEE}%
\thanks{%
     D. S. Krotov is with the Sobolev Institute of Mathematics,
     pr. Akademika Koptyuga 4,
     Novosibirsk 630090, Russia (e-mail: krotov@math.nsc.ru),
\and
     and also with the Novosibirsk State University,
     Pirogova 2, Novosibirsk 630090, Russia}
}

\date{}
\maketitle
\begin{abstract}
 In this paper,
two different Gray-like maps
from $Z_p^\alpha\times Z_{p^k}^\beta$, where $p$ is prime,
to $Z_p^n$, $n={\alpha+\beta p^{k-1}}$, denoted by $\varphi$ and $\Phi$, respectively, are presented.
We have determined the connection between the weight enumerators among the image codes under these two mappings.
We show that if $C$ is a $Z_p Z_{p^k}$-additive code, and $C^\bot$ is its dual,
then the weight enumerators of the image $p$-ary codes $\varphi(C)$ and $\Phi(C^\bot)$ are formally dual.
This is a partial generalization of
[D. S. Krotov, On $Z_{2^k}$-dual binary codes,
IEEE Trans. Inform. Theory 53 (2007), 1532--1537, arXiv:\href{http://arXiv.org/abs/math/0509325}{math/0509325}],
and the result is generalized to odd characteristic $p$
and mixed alphabet. 
Additionally, a construction of $1$-perfect additive codes
in the mixed $Z_p Z_{p^2}\ldots Z_{p^k}$ alphabet 
is given.
\end{abstract}


\section{Introduction}

Quaternary codes have attracted people's attention since the 80th, due to their relationship to some well-known nonlinear binary codes \cite{HammonsOth:Z4_linearity, Nechaev82,Nechaev:Kerdock_cyclic_form}. It was shown \cite{HammonsOth:Z4_linearity} that some good nonlinear codes, including the Kerdock codes,
can be viewed as the image codes of $\Z_4$ cyclic codes under the so called Gray mapping.
In \cite{Carlet:Z2k-linear}, the Gray map was generalized to construct new $\Z_{2^k}$-linear codes, such as the generalized Kerdock codes and Delsarte-Goethals codes. The definition of the generalized Gray map can be found in Section~\ref{s:prel}, i.e., $\varphi_k$ for the special case $p=2$.
However, it is worth noting that if $\mathcal{C}$ is a linear code over $\Z_{2^k}$,
then it can be proved that the weight enumerators of $\varphi_k(\mathcal{C})$ and $\varphi_k(\mathcal{C}^\bot)$ do not in general satisfy the MacWilliams identity,
i.e., are not formally dual.
On the positive side, in \cite{Kro:2007Z2k}, it was introduced another generalization of the Gray map, which can be viewed as dual to  $\varphi_k$ in some sense. It was shown that
the weight distributions of the image codes under these two generalized Gray mappings satisfy the MacWilliams identity.

Additive codes (mixed alphabet codes) were first defined by Delsarte in 1973 in terms of association schemes \cite{Delsarte:1973}. In the following 1997, the translation invariant propelinear codes were first introduced by Rif\`a and Pujol \cite{RifPuj:1997}. As follows from Delsarte's results,
 any abelian binary propelinear codes has the form $\Z_2^\gamma\times \Z_4^\delta$ for some nonnegative integers $\gamma$ and $\delta$. In general, a $\Z_2\Z_4$-additive code
is defined to be a subgroup of $\Z_2^\alpha \times\Z_4^\beta$; this is a generalization of the usual binary linear codes and quaternary linear codes.
Later, the structure and properties of $\Z_2\Z_4$-additive codes have been intensely studied, including generator matrix, duality, kernel and rank, see \cite{BDF:self-dual:2012, BFPRV2010,BorRif:1999,FePuVi08rk}.
In~\cite{RRR:2011}, perfect $\Z_2\Z_4$-additive codes were shown to be potentially useful in the steganography, for hiding information;
this application can be considered as a partial motivation for further researches in the area of additive codes, including our current results.
Furthermore, $\Z_2\Z_4$-additive codes were generalized to $\Z_2\Z_{2^s}$-additive codes in \cite{AydSiap:2013}, these codes are meaningful because they provided good binary codes via Gray maps. Then, the structure and the duality of $\Z_{p^r}\Z_{p^s}$-additive codes were discussed in \cite{AydSiap:2015}. Note that the last two papers mentioned considered additive codes in the Lee metric space. We study another generalization of the $\Z_2\Z_4$-additive codes, related to the homogeneous metric and the metric that can be considered as dual to homogeneous.


In the present paper, we introduce two generalized Gray maps on $\Z_p\Z_{p^k}$ (see Section~\ref{s:prel}, the definitions of $\varphi$ and $\Phi$), and prove that if $\mathcal{C}$ is a $\Z_p\Z_{p^k}$-additive code and $\mathcal{C}^\bot$ is its dual, then the images $\varphi(\mathcal{C})$ and $\Phi(\mathcal{C}^\bot)$ are formally dual, i.e., satisfy the MacWilliams identity. The result has a nature extension to the mixed $\Z_p \Z_{p^2}\dots\Z_{p^k}$ alphabet, see Remark~\ref{remark1}.

It should be mentioned that, while the $Z_2Z_4$-linear
binary codes are known as a partial case of the propelinear codes \cite{RifPuj:1997},
in  general  the $p$-ary codes
obtained by the Gray map from $\Z_p\Z_{p^k}$-additive codes are not proven
to be propelinear even for the case $k=2$.
The study of this question is an interesting topic for futher research.

The manuscript is organized as follows. Section~\ref{s:prel} fixes some notations and definitions for this paper, we introduce two generalized Gray-like maps on $\Z_p\Z_{p^k}$. In addition, we describe the connection between these two mappings. The main results are given in Section~\ref{s:MW}, where we establish the MacWilliams identity between the image codes $\varphi(\mathcal{C})$ and $\Phi(\mathcal{C}^\bot)$ (Theorem \ref{main}). Section~\ref{s:perfect} gives a construction of $1$-perfect additive codes in the mixed $\Z_p \Z_{p^2}\dots\Z_{p^k}$ alphabet with a special distance, such that the $\Phi$-image is a perfect code over $\Z_p$.

\section{Preliminaries}\label{s:prel}
\subsection{Linear Codes}\label{ss:linear}

Let $p$ be an odd prime number. Denote by $\Z_p$ and $\Z_{p^k}$ the rings of integers modulo $p$ and $p^k$, respectively.
The \emph{Hamming weight} of $\mathbf x=(x_1,\ldots,x_n)\in \Z_p^n,$ denoted by $\mathrm{wt}(\mathbf x)$, is the number of indices $i$ where $x_i \neq 0$.
A linear code $C$ of length $n$ over the ring $\Z_p$ is a $\Z_p$-submodule of $\Z_p^n$.
If the cardinality of the code $C$ is $M$, and its \emph{(minimum) distance}, denoted by $d$, is defined as the minimum Hamming weight of its nonzero elements, then it is sometimes referred to as an $(n,M,d)$ code over $\Z_p$.
The \emph{weight enumerator} of the code $C$ is defined by $W_C(X,Y)=\sum\limits_{\textbf{c}\in C}X^{\mathrm{wt}(\textbf{c})}Y^{n-\mathrm{wt}(\textbf{c})}$, a homogeneous polynomial of degree $n$ in two variables.
From now on, we will focus on $\mathbb{Z}_p\mathbb{Z}_{p^k}$-additive codes. A \emph{$\mathbb{Z}_p\mathbb{Z}_{p^k}$-additive code} is a $\mathbb{Z}_{p^k}$-submodule of $\mathbb{Z}_p^\alpha \times \mathbb{Z}_{p^k}^\beta$.
Throughout this paper we use calligraphic symbols like $\mathcal{C}$ to denote codes in the mixed $\Z_p\Z_{p^k}$ alphabet (even if the $\Z_p$ part is empty), and
we use standard symbols like $C$ to denote codes over $\Z_{p}$.

Let $\mathbf{c}=(\mathbf{v}|\mathbf{w})\in \mathbb{Z}_p^\alpha \times \Z_{p^k}^\beta$.
For $l\in \Z_{p^k}$, where $l$ can be expressed as $l=l_0+l_1p+\dots+l_{k-1}p^{k-1}$, and $0\leq l_i\leq p-1$, we have
$$l\mathbf{c}=l(\mathbf{v}|\mathbf{w})=(l_0\mathbf{v}|l\mathbf{w}).$$
The \emph{inner product} between $(\mathbf{v}_1|\mathbf{w}_1)$ and $(\mathbf{v}_2|\mathbf{w}_2)$ in $\mathbb{Z}_p^\alpha \times \Z_{p^k}^\beta$ can be written as follows:
$$\langle(\mathbf{v}_1|\mathbf{w}_1),(\mathbf{v}_2|\mathbf{w}_2)\rangle=p^{k-1}\langle \mathbf{v}_1,\mathbf{v}_2\rangle + \langle \mathbf{w}_1,\mathbf{w}_2\rangle \in \Z_{p^k},$$
where $\langle \cdot , \cdot \rangle$ denotes the usual inner product
$$ \langle(x_0,\ldots, x_{n-1}),(y_0,\ldots, y_{n-1})\rangle = x_0y_0+\ldots+x_{n-1}y_{n-1}. $$
Note that the result of the inner product $\langle \mathbf{v}_1,\mathbf{v}_2\rangle$ is from $\Z_p$,
and multiplication of its value by $p^{k-1}$ should be formally understood as the natural homomorphism from $\Z_p$ into $\Z_{p^k}$.
The \emph{dual code} $\mathcal{C}^\bot$ of a $\mathbb{Z}_p\Z_{p^k}$-additive code $\mathcal{C}$ is defined in the standard way by
\begin{IEEEeqnarray*}{rl}
 \mathcal{C}^\bot=\big\{(\mathbf{x}|\mathbf{y})
 & {}\in \mathbb{Z}_p^\alpha \times \Z_{p^k}^\beta:
 \\ &
 {
 \langle(\mathbf{x}|\mathbf{y}),(\mathbf{v}|\mathbf{w})\rangle=0 ~\rm {for ~all}~ (\mathbf{v}|\mathbf{w})\in \mathcal{C}\big\}.}
\end{IEEEeqnarray*}
Readily, the dual code is also a $\mathbb{Z}_p\Z_{p^k}$-additive code.

\subsection{Gray Maps}
In this subsection, we will introduce two different Gray-like functions $\varphi$ and $\Phi$ from $\mathbb{Z}_p^\alpha \times \Z_{p^k}^\beta$ to $\Z_p^n$, $n={\alpha+\beta p^{k-1}}$.
The first Gray-like map $\varphi$ corresponds to the homogeneous metric over $\Z_{p^k}$,
and the second function $\Phi$ can be regarded as the dual of the first case
(in general, $\Phi$ is a multi-valued function; so, formally it is a map
from $\mathbb{Z}_p^\alpha \times \Z_{p^k}^\beta$ to
the set of subsets of $\Z_p^n$).
More details are given as follows.

Let $P$ be the linear code over $\Z_p$ with the generator matrix $A$ in the form $\binom{\bar{1}}{B}$, where $\bar{1}$ denotes the all-$1$ vector of length $p^{k-1}$ and the columns of the matrix $B$ are all different vectors in $\Z_{p}^{k-1}$. Then $P$ is a linear two-weight code of size $p^k$ with nonzero weights $(p-1)p^{k-2}$ and $p^{k-1}$. Arrange the codewords in  $P=\{\mathbf{c}_0,\mathbf{c}_1,\ldots,\mathbf{c}_{p^{k}-1}\}$ in such a way that $\mathbf{c}_0=(0,\ldots,0)$  and for all $i$, $0\leq i\leq p^k-1$, and $j$, $0\leq j\leq p-1$, the codeword $\mathbf{c}_{i+jp^{k-1}}-\mathbf{c}_i$ has the form   $(j,j,\ldots,j)$.

\begin{example}
Let $p=k=3$. From the description above, the generator matrix $A$ has the form as follows:
$$\left(
  \begin{array}{ccccccccc}
    1 & 1 & 1 & 1 & 1 & 1 & 1 & 1 & 1 \\
    0 & 0 & 0 & 1 & 1 & 1 & 2 & 2 & 2 \\
    0 & 1 & 2 & 0 & 1 & 2 & 0 & 1 & 2 \\
  \end{array}
\right)$$
and the weight enumerator of the code $P$ with the generator matrix $A$ is given by
$$W_C(X,Y)=X^9 + 24X^3Y^6 + 2Y^9.$$
\end{example}

\begin{definition}
Define the \emph{Gray map $\varphi_k$} from $\Z_{p^k}^n$ to $\Z_p^{p^{k-1}n}$ as
$$\varphi_k(x_1,x_2,\ldots,x_n)=(\mathbf{c}_{x_1},\mathbf{c}_{x_2},\ldots,\mathbf{c}_{x_n}),$$
where $x_i\in \Z_{p^k}$, $\mathbf{c}_{x_i}\in \Z_p^{p^{k-1}}$ and $1 \leq i \leq n$.
\end{definition}
Then the weight function $\mathrm{wt}^*$ is defined by:
\begin{equation*}
\mathrm{wt}^*(x) = \left\{
\begin{array}{rl}
0  &  \text{if } x = 0,\\
p^{k-1}   & \text{if } x \in p^{k-1}\Z_{p^k}\backslash \{0\}, \\
(p-1)p^{k-2}  & \text{otherwise}.
\end{array} \right.
\end{equation*}
The definition of the weight function here is consistent with the homogeneous metric introduced in \cite{ConstHeise}, and we know this weight function can be expressed by a character sum \cite{Honold:2001}. The corresponding distance $d^*$ on $\Z_{p^k}^n$ is defined as follows:
$$d^*(\mathbf{x},\mathbf{y})=\sum_{i=1}^n\mathrm{wt}^*(y_i-x_i),$$
where $\mathbf{x}=(x_1,x_2,\ldots,x_n),\ \mathbf{y}=(y_1,y_2,\ldots,y_n)\in \Z_{p^k}^n$.

It is easy to check that the mapping $\varphi_k$ is an isometric embedding of $(\Z_{p^k}^n, d^*)$ into $(\Z_p^{p^{k-1}n},d_H)$, where $d_H$ denotes the usual Hamming distance. If $\mathcal C$ is a code with parameters $(n,M,d^*)$ over $\Z_{p^k}$, then the image code $C=\varphi_k({\mathcal C})$ is a code with parameters $(p^{k-1}n,M,d_H)$ over $\Z_p$.
Then, we define the Gray-like map $\varphi$ for elements $(\mathbf{v}|\mathbf{w})\in \mathbb{Z}_p^\alpha \times \Z_{p^k}^\beta$, by $\varphi((\mathbf{v}|\mathbf{w}))=(\mathbf{v}|\varphi_k(\mathbf{w})).$

Next, we introduce another Gray map from its dual side.
Let the code $D$ be the dual to the linear code $P$ with the parity-check matrix $A$ introduced above.
If $p$ is odd, then any two columns of $A$ are linearly independent and $A$ has three columns that are linearly dependent;
 so, the dual code $D$ is a linear code with parameters $(p^{k-1},p^{p^{k-1}-k},3)$ over $\Z_p$.
 If $p=2$, then $D$ is a binary $(2^{k-1},2^{2^{k-1}-k},4)$ extended Hamming code.
 Consider all the cosets of the linear code $D$, write as $D_i$ for $i=0,1,\ldots,p^k-1$. Then the set $\{D=D_0,D_1,\ldots, D_{p^k-1}\}$ forms a partition of $\Z_p^{p^{k-1}}$.
 Additionally, we require the sum of all coordinates of a codeword of $D_i$ to be equal to $i\bmod p$ (for every coset, this sum is a constant, because $(1,\ldots,1)$ is a row of $A$; so, this condition can be satisfied by an appropriate enumeration of the cosets).

\begin{definition}
Define the \emph{Gray map $\Phi_k$} from $\Z_{p^k}^n$
to the set of subsets of $\Z_p^{p^{k-1}n}$ by
$$\Phi_k\big((x_1,x_2,\ldots,x_n)\big)=D_{x_1}\times D_{x_2}\times \dots \times D_{x_n},$$
where $x_i\in \Z_{p^k}$, $D_0$, $D_1$, \ldots, $D_{p^k-1}$
are the cosets of $D$, ordered as described above.
\end{definition}

For $x\in \Z_{p^k}$, the weight function $\mathrm{wt}^\diamond$ is defined as follows:
\begin{equation*}
\mathrm{wt}^\diamond(x) = \left\{
\begin{array}{rl}
0  &  \text{if } x = 0,\\
1   & \text{if } p\nmid x , \\
2  &  \text{if } p|x,\ \text{and} \ x\neq 0.
\end{array} \right.
\end{equation*}
The corresponding distance $d^\diamond$ on $\Z_{p^k}^n$ is defined as follows:
$$d^\diamond(\mathbf{x},\mathbf{y})=\sum_{i=1}^n\mathrm{wt}^\diamond(y_i-x_i),$$
where $\mathbf{x}=(x_1,x_2,\ldots,x_n),\ \mathbf{y}=(y_1,y_2,\ldots,y_n)\in \Z_{p^k}^n$.

The map $\Phi_k$
introduced here is not an isometric embedding of
$(\Z_{p^k}^n, d^\diamond)$
into $(\Z_p^{p^{k-1}n},d_H)$,
but still carry some partial isometric properties.
If a code ${\mathcal C}$ with parameters
$(n,|{\mathcal C}|,d^\diamond)$
over $\Z_{p^k}$,
then the image code
$$\Phi_k({\mathcal C}) = \bigcup_{\mathbf{c}\in \mathcal C} \Phi_k(\mathbf{c})$$
is a code with parameters
$(p^{k-1}n,|{\mathcal C}|\cdot
p^{(p^{k-1}-k)n},d^\prime)$
over $\Z_p$,
where $d^\prime=\min\{3,d^\diamond\}$ for odd $p$ and  $d^\prime=\min\{4,d^\diamond\}$ in the case $p=2$.

Similarly, we define the Gray-like map $\Phi$
for the elements $(\mathbf{v}|\mathbf{w})$ of $\mathbb{Z}_p^\alpha \times \Z_{p^k}^\beta$
and for the subsets ${\mathcal C}$  of $\mathbb{Z}_p^\alpha \times \Z_{p^k}^\beta$:
$$\Phi((\mathbf{v}|\mathbf{w}))
=
\{\mathbf{v}\} \times \Phi_k(\mathbf{w}),
\qquad
\Phi({\mathcal C}) = \bigcup_{(\mathbf{v}|\mathbf{w})\in \mathcal C} \Phi\big((\mathbf{v}|\mathbf{w})\big).$$

\section{$\mathbb{Z}_{p}$-duality for image codes}\label{s:MW}

In this section, we determine the weight relationship between the image codes $\varphi(\mathcal{C})$ and $\Phi(\mathcal{C}^\perp)$ under the maps $\varphi$ and $\Phi$, respectively. For this purpose, we have to start with the weight distributions of the cosets of the linear code $D$. We first observe that $D$ satisfies the hypothesis of the following proposition.

\begin{proposition}\label{Delsarte}
Let $d$ be the Hamming distance of a code $C$, and let $s$ be the number of nonzero Hamming weights of its dual.
\begin{enumerate}
    \item[\rm (i)] {\rm \cite[Theorem~5.13]{Delsarte:1973}}
    If $2s-1\le d \le 2s+1$,
then the weight distribution
of a coset of $C$
depends only on its minimum weight.
    \item[\rm (ii)] {\rm \cite[Theorem~5.10]{Delsarte:1973}}
    The minimum weight of a coset of $C$ cannot be larger than $s$.
\end{enumerate}
\end{proposition}
Therefore, the cosets of $D$ have only $3$ different weight distributions.
Trivially, the coset $D_0=D$ has a word of weight $0$. All other cosets $D_i$, $p|i$, have minimum weight $2$, because their words are orthogonal to the all-one word and by Proposition~\ref{Delsarte}(ii).
The remaining cosets $D_i$, $p\nmid i$, have minimum weight $1$, because their number coincides with the number of weight-$1$ words.

\begin{lemma}\label{1}
Let the linear code $D$ be introduced in Section~\ref{s:prel}, and let $\{D=D_0,D_1,\ldots, D_{p^k-1}\}$ be the coset partition of $\Z_p^{p^k-1}$ introduced above. Then
\begin{itemize}
  \item the weight enumarator od $D_0$ satisfies the identity
  \begin{IEEEeqnarray*}{rCl}
  \IEEEeqnarraymulticol{3}{l}
    {\frac{1}{|D_0|}W_{D_0}(X+(p-1)Y,X-Y)} \\
   \ &=&X^{p^{k-1}}\!\!+(p-1)Y^{p^{k-1}}
     \!\!+(p^k-p)X^{p^{k-2}}Y^{(p-1)p^{k-2}};
  \end{IEEEeqnarray*}
  \item if $i\equiv 0 \bmod p$ and $i\neq 0$, then
  \begin{multline}\nonumber
    \frac{1}{|D_i|}W_{D_i}(X+(p-1)Y,X-Y) \\
    {}=\,X^{p^{k-1}}\!\!+(p-1)Y^{p^{k-1}}
     \!\!-pX^{p^{k-2}}Y^{(p-1)p^{k-2}};
  \end{multline}
  \item if $i\not\equiv 0 \bmod p$, then
  $$\frac{1}{|D_i|}W_{D_i}(X+(p-1)Y,X-Y)=X^{p^{k-1}}-Y^{p^{k-1}}.$$
\end{itemize}
\end{lemma}

\begin{proof}
The weight enumerator of $D_0$ is easy to obtain from the MacWilliams identity, since its dual $P$ is a two-weight code. For the rest cases, let $H$ be a linear code of length $p^{k-1}$ with the generator matrix containing the all-1 vector over $\Z_p$ as the only row. Denote the dual of $H$ by $H^\bot$. We know that the weight enumerator of $H^\bot$ is
\begin{multline} \nonumber
  W_{H^\bot}(X,Y)
   = \frac{1}{|H|}W_{H}(X+(p-1)Y,X-Y) \\
   {}= \frac{1}{p}\big((X+(p-1)Y)^{p^{k-1}}+(p-1)(X-Y)^{p^{k-1}}\big).
\end{multline}
On the other hand, we know
\begin{IEEEeqnarray*}{rCL}
  \IEEEeqnarraymulticol{3}{l}{ W_{D_0}(X,Y) }
  \\
  \quad&=  \frac{1}{p^k}&\Big(\big(X+(p-1)Y\big)^{p^{k-1}}\!\!+(p-1)(X-Y)^{p^{k-1}} \\
    &&+(p^k-p)\big(X+(p-1)Y\big)^{p^{k-2}}(X-Y)^{(p-1)p^{k-2}}\Big),
\end{IEEEeqnarray*}
and $H^\bot=D_0\cup \bigcup\limits_{i:\,p|i,\,i\neq 0}D_i$; here all $D_i$'s have the same enumerators. Then we have
\begin{multline} \nonumber
  W_{D_i}(X,Y) = \frac{1}{p^{k-1}-1}\big(W_{H^\bot}(X,Y)-W_{D_0}(X,Y)\big) \\
   = \frac{1}{p^k}\Big(\big(X+(p-1)Y\big)^{p^{k-1}}\!\!+(p-1)(X-Y)^{p^{k-1}}\\
   {}-p\big(X+(p-1)Y\big)^{p^{k-2}}(X-Y)^{(p-1)p^{k-2}}\Big).
\end{multline}
Then, by a simple variable substitutions, we obtain the result directly.
The remaining case follows from the fact that
$\Z_p^{p^{k-1}}\backslash H^\bot=\bigcup\limits_{i:\,p\nmid i}D_i$.
\end{proof}

The \emph{complete weight enumerator} of a $\Z_p\Z_{p^k}$-additive code $\mathcal{C}$ is
\begin{IEEEeqnarray*}{r}
W_\mathcal{C}\big(X_i,Y_j;\ i=0,1,\ldots,p-1,\ j=0,1,\ldots,p^k-1\big)
\\
=\sum_{(\mathbf{v}|\mathbf{w})\in \mathcal{C}}\bigg(\prod_{i=1}^\alpha X_{v_i}\bigg)\bigg(\prod_{j=1}^\beta Y_{w_j}\bigg),
\\
(\mathbf{v}|\mathbf{w})=(v_1,\ldots,v_{\alpha}|w_1,\ldots,w_{\beta}).
\end{IEEEeqnarray*}
Then, we define the polynomial
$SW_\mathcal{C}(X,S,Y,Z,T)$ obtained from $W_\mathcal{C}$
by 
\begin{itemize}
 \item substituting $X$ for $X_0$, $Y$ for $Y_0$, 
 \item  identifying $S$ and all $X_i$'s with $i\neq 0$,
 \item identifying $Z$ and all $Y_j$'s such that $p \nmid j$, and
 \item identifying $T$ and all $Y_j$'s such that $p \mid j$ and $j\neq 0$.
\end{itemize}

Let $\omega_1$ and $\omega_2$ be the complex numbers $e^{2\pi i/p}$ and $e^{2\pi i/p^k}$,
respectively.
For a complex-valued function $f$ defined on $\Z_p^\alpha \times \Z_{p^k}^\beta$, denote
$$
\widehat{f}(\mathbf{z})=
\sum\limits_{\mathbf{u}\in \Z_p^\alpha\times \Z_{p^k}^\beta}
     \omega_2^{\langle \mathbf{z},\mathbf{u}\rangle}f(\mathbf{u}),
\qquad
\mathbf{z}\in \Z_p^\alpha \times \Z_{p^k}^\beta
$$
Here $\omega_2^{\langle (\mathbf{z}',\mathbf{z}''),(\mathbf{u}',\mathbf{u}'')\rangle}$ can be written as $\omega_1^{\langle \mathbf{u}',\mathbf{z}'\rangle}\omega_2^{\langle \mathbf{u}'',\mathbf{z}''\rangle}$, where $\mathbf{u}',\mathbf{z}'\in \Z_p^\alpha$ and $\mathbf{u}'',\mathbf{z}''\in \Z_{p^k}^\beta$.
The function $\widehat{f}$ is called the \emph{Fourier transform} of $f$.

\begin{lemma}\label{hadmard}
Let $\mathcal{C}$ be an additive code in $\Z_p^\alpha\times \Z_{p^k}^\beta$, and let $\mathcal{C}^\bot$ be its dual. Then for every complex-valued function $f$ on $\Z_p^\alpha \times \Z_{p^k}^\beta$,
$$\sum\limits_{\mathbf{z}\in \mathcal{C}^\bot}f(\mathbf{z})=\frac{1}{|\mathcal{C}|}\sum\limits_{\mathbf{u}\in \mathcal{C}}\widehat{f}(\mathbf{u}).$$
\end{lemma}
\begin{proof}
We have
\begin{multline} \nonumber
  \sum\limits_{\mathbf{u}\in \mathcal{C}}\widehat{f}(\mathbf{u}) = \sum\limits_{\mathbf{u}\in \mathcal{C}}\sum\limits_{\mathbf{z}\in \Z_p^\alpha\times \Z_{p^k}^\beta}\omega_2^{\langle\mathbf{u},\mathbf{z}\rangle}f(\mathbf{z})\\
  = \sum\limits_{\mathbf{z}\in \Z_p^\alpha\times \Z_{p^k}^\beta}f(\mathbf{z}) \sum\limits_{\mathbf{u}\in \mathcal{C}}\omega_2^{\langle\mathbf{u},\mathbf{z}\rangle}.
  \end{multline}
If $\mathbf{z} \in \mathcal{C}^\bot$, then $\langle\mathbf{u},\mathbf{z}\rangle=0$ for all $\mathbf{u}\in \mathcal{C}$. Hence, the inner sum $\sum\limits_{\mathbf{u}\in \mathcal{C}}\omega_2^{\langle\mathbf{u},\mathbf{z}\rangle}$ is equal to $|\mathcal{C}|$.
On the other hand, if $\mathbf{z} \notin \mathcal{C}^\bot$, then there exits $\mathbf{u}_0\in \mathcal{C}$ such that $\langle\mathbf{u}_0,\mathbf{z}\rangle=\lambda \neq 0$. Since $\mathcal{C}$ is an $\Z_p\Z_{p^k}$-additive code, for the inner sum $\sum\limits_{\mathbf{u}\in \mathcal{C}}\omega_2^{\langle\mathbf{u},\mathbf{z}\rangle}$, we have
\begin{multline}\nonumber
  \sum\limits_{\mathbf{u}\in \mathcal{C}}\omega_2^{\langle\mathbf{u},\mathbf{z}\rangle} = \sum\limits_{\mathbf{u}\in \mathcal{C}}\omega_2^{\langle\mathbf{u}+\mathbf{u}_0,\mathbf{z}\rangle}
  = \omega_2^{\langle\mathbf{u}_0,\mathbf{z}\rangle} \sum\limits_{\mathbf{u}\in \mathcal{C}}\omega_2^{\langle\mathbf{u},\mathbf{z}\rangle}
\end{multline}
Since $\omega_2^{\langle\mathbf{u}_0,\mathbf{z}\rangle}\neq 1$, the inner sum is zero. Therefore,
$$\sum\limits_{\mathbf{u}\in \mathcal{C}}\widehat{f}(\mathbf{u})=|\mathcal{C}|\sum\limits_{\mathbf{z}\in \mathcal{C}^\bot}f(\mathbf{z}).$$
Then the result follows.
\end{proof}

Let $\mathbf{z}=(\mathbf{z}',\mathbf{z}'')\in \Z_p^\alpha\times \Z_{p^k}^\beta$ and $$ f(\mathbf{z})=\prod\limits_{i=0}^{p-1}X_i^{w_i(\mathbf{z}')}\prod\limits_{j=0}^{p^k-1}Y_j^{w_j(\mathbf{z}'')},$$ where $w_i(\mathbf{z}')$ (respectively, $w_j(\mathbf{z}'')$) denotes the number of $i$ (respectively, $j$) in $\mathbf{z}'$ (respectively, $\mathbf{z}''$).
By computing the Fourier transform $\widehat{f}(\mathbf{z})$ of $f(\mathbf{z})$, and according to Lemma \ref{hadmard},
we find that the complete weight enumerator $W_{\mathcal{C}^\bot}$ satisfies the MacWilliams identity
\begin{IEEEeqnarray*}{rCr}
\IEEEeqnarraymulticol{3}{l} {
W_{\mathcal{C}^\bot}\big(X_i,Y_j;i=0,1,\ldots,p-1,\ j=0,1,\ldots,p^k-1\big) }
\\
\quad&=&\frac{1}{|\mathcal{C}|}W_\mathcal{C}\bigg(\sum_{t=0}^{p-1}\omega_1^{it}X_t,\sum_{s=0}^{p^k-1}\omega_2^{js}Y_s;\ i=0,1,\ldots,p-1, \quad
\\
&& j=0,1,\ldots,p^k-1\bigg).
\end{IEEEeqnarray*}
If $C=\varphi(\mathcal{C})$, then we can obtain the weight enumerator of $C$ from the complete weight enumerator of $\mathcal{C}$ by
\begin{itemize}
  \item replacing $X_0$ by $X$,
  \item replacing  $Y_0$ by $X^{p^{k-1}}$,
   \item replacing $X_i$ by $Y$ \ for $i\neq 0$,
  \item replacing $Y_i$ by $Y^{p^{k-1}}$ \ if $p^{k-1}|i$ and $i\neq 0$,
  \item replacing $Y_i$ by $X^{p^{k-2}}Y^{(p-1)p^{k-2}}$ \ if $p^{k-1}\nmid i$.
\end{itemize}

Now, for each $j\in \{0,1,\ldots,p^k-1\}$, we start to compute the value of the expression $\sum\limits_{s=0}^{p^k-1}\omega_2^{js}Y_s$ after phrase transformations as follows:
\begin{itemize}
 \item If $j=0$, we have
 $$X^{p^{k-1}}\!\!+(p-1)Y^{p^{k-1}}\!\!+(p^k-p)X^{p^{k-2}}Y^{(p-1)p^{k-2}}.$$
 \item If $p\mid j$ and $j\neq 0$,
 then each coefficient at $Y_s$, $s \equiv 0 \bmod p^{k-1}$, is equal to $1$,
 and the sum of the remaining coefficients is $-p$. Hence, we get
     $$X^{p^{k-1}}\!\!+(p-1)Y^{p^{k-1}}\!\!-pX^{p^{k-2}}Y^{(p-1)p^{k-2}}.$$
 \item If $p\nmid j$, then it is easy to check that the sum of the coefficients of $Y_s$ for $p^{k-1}\mid s$ and $i\neq 0$ is equal to $-1$; the rest follows from the equality $\sum\limits_{s=0}^{p^k-1}\omega_2^{js}=0$. We get $$X^{p^{k-1}}\!\!-Y^{p^{k-1}}.$$
 \end{itemize}
Similarly, the value of the expression  $\sum\limits_{t=0}^{p-1}\omega_1^{it}X_t$, after phrase transformations, is as follows:
\begin{itemize}
 \item if $i=0$, we have $X+(p-1)Y$;
 \item if $i\neq 0$, we have $X-Y$.
 \end{itemize}

From the discussion above, we deduce the following lemma.
\begin{lemma}\label{2}
Let $\mathcal{C}$ be a $\Z_p\Z_{p^k}$-additive code,
and let ${C}_\bot=\varphi(\mathcal{C}^\bot)$; then we have
\begin{IEEEeqnarray*}{Rl}
  W_{{C}_\bot}(&X,Y) =
  \frac{1}{|\mathcal{C}|}
  SW_{\mathcal{C}}\bigg(
  X+(p-1)Y,\ X-Y, \\
  & X^{p^{k-1}}+(p-1)Y^{p^{k-1}}
  +(p^k-p)X^{p^{k-2}}Y^{(p-1)p^{k-2}},\\
  & X^{p^{k-1}}-Y^{p^{k-1}},\\
  & X^{p^{k-1}}+(p-1)Y^{p^{k-1}}-pX^{p^{k-2}}Y^{(p-1)p^{k-2}}\bigg).
\end{IEEEeqnarray*}
\end{lemma}

Next, we consider the connection between the weight enumerators $W_{\Phi(\mathcal{C})}$ and $SW_{\mathcal{C}}$.
\begin{lemma}\label{3}
Let $\mathcal{C}$ be an additive code in $\Z_p^\alpha\times \Z_{p^k}^\beta$, and ${\tilde{C}}=\Phi(\mathcal{C})$. Then we have
\begin{multline}\nonumber
W_{{\tilde{C}}}(X,Y)=SW_{\mathcal{C}}\big(X,\,Y,\,W_{D_0}(X,Y),\\
W_{D_1}(X,Y),\,W_{D_p}(X,Y)\big),
\end{multline}
where $\{D_0,D_1,\ldots,D_{p^k-1}\}$ is the partition of $\Z_p^{p^k-1}$ defined above.
\end{lemma}
\begin{proof}
Every codeword $\mathbf{c}=(x_1,x_2,\ldots,x_\alpha|y_1,y_2,\ldots, y_\beta)$ of $\mathcal{C}$
contributes $SW_{x_1}\ldots SW_{x_\alpha}SW'_{y_1}\ldots SW'_{y_\beta}$ to $SW_\mathcal{C}(X,S,Y,Z,T)$,
where
\begin{itemize}
 \item $SW_0$ equals $X$,
\item $SW_i$ equals $S$ for $i\neq 0$,
\item $SW'_0$ equals $Y$,
\item $SW'_i$ equals $Z$ if $p \nmid i$, and
\item $SW'_i$ equals $T$ if $p \mid i$ and $i\neq 0$.
\end{itemize}

On the other hand, according to the definition of the map $\Phi$ introduced in Section~\ref{s:prel}, the image $\Phi(\mathbf{c})$ contributes $SW_{x_1}\dots SW_{x_\alpha}W_{D_{y_1}}(X,Y)\dots W_{D_{y_\beta}}(X,Y)$ to $W_{{C}}(X,Y)$. From Lemma \ref{1}, we know that $W_{D_{i}}(X,Y)$ equals $W_{D_1}(X,Y)$ if $p\nmid i$, and $W_{D_{i}}(X,Y)$ equals $W_{D_p}(X,Y)$ if $p|i$ and $i\neq 0$. The desired result follows.
\end{proof}

Now, we introduce the main result of this section.
\begin{theorem}\label{main}
Let $\mathcal{C}$ be an additive code in $\Z_p^\alpha\times \Z_{p^k}^\beta$, and let $\mathcal{C}^\bot$ be the dual of  $\mathcal{C}$.
Denote ${C}=\varphi(\mathcal{C})$ and ${{C}}_\bot=\Phi(\mathcal{C}^\bot)$. Then we have $$W_{{C}}(X,Y)=\frac{1}{|{C}_\bot|}W_{{{C}}_\bot}(X+(p-1)Y,X-Y).$$
\end{theorem}
\begin{proof}
From Lemmas \ref{1} and \ref{2}, we have
\begin{IEEEeqnarray*}{Rl}
  W_{{C}}(X,Y) =\frac{1}{|\mathcal{C}^\bot|}
   SW_{\mathcal{C}^\bot}&\Big(X+(p-1)Y,\ X-Y,\
  \\
  &\frac{1}{|D_0|}W_{D_0}(X+(p-1)Y,X-Y),
  \\
  &\frac{1}{|D_1|}W_{D_1}(X+(p-1)Y,X-Y),
  \\
  &\frac{1}{|D_p|}W_{D_p}(X+(p-1)Y,X-Y)\Big)\\
   =       \frac{1}{|\mathcal{C}^\bot|}
   \bigg(\frac{p^k}{p^{p^{k-1}}}\bigg)^\beta SW_{\mathcal{C}^\bot}
   &\Big(X+(p-1)Y,\ X-Y,\\
   &\ \  W_{D_0}(X+(p-1)Y,X-Y),\\
   &\ \  W_{D_1}(X+(p-1)Y,X-Y),\\
   &\ \  W_{D_p}(X+(p-1)Y,X-Y)\Big).
\end{IEEEeqnarray*}
We know that $|\mathcal{C}^\bot|\bigg(\frac{p^{p^{k-1}}}{p^{k}}\bigg)^\beta=|{{C}}_\bot|$, and according to Lemma \ref{3}, we have
\begin{multline}\nonumber
 SW_{\mathcal{C}^\bot}\Big(X+(p-1)Y,\ X-Y,
 \\  W_{D_0}\big(X+(p-1)Y,X-Y\big), \\
  W_{D_1}\big(X+(p-1)Y,X-Y\big),\\
  W_{D_p}\big(X+(p-1)Y,X-Y\big)\Big)\\
 {}= W_{{{C}}_\bot}(X+(p-1)Y,X-Y).
\end{multline}
This completes the proof.
\end{proof}

\begin{remark}\label{remark1}
Theorem \ref{main} can be extended to more general case. Let $\mathcal{C}$ be an additive code in $\Z_p^{\alpha_1} \times \Z_{p^2}^{\alpha_2} \times \dots \times \Z_{p^k}^{\alpha_k}$, and $\mathcal{C}^\bot$ be its dual. Let ${C}=\varphi(\mathcal{C})$ and ${C}_\bot=\Phi(\mathcal{C}^\bot)$. Then we have $$W_{{C}}(X,Y)=\frac{1}{|{C}_\bot|}W_{{C}_\bot}(X+(p-1)Y,X-Y).$$
Note that, for given
$\mathbf{v} = (\mathbf{v}_1|\mathbf{v}_2|\ldots|\mathbf{v}_k)$ and
$\mathbf{u} = (\mathbf{u}_1|\mathbf{u}_2|\ldots|\mathbf{u}_k)$ from
$\mathcal{C}\subseteq \Z_p^{\alpha_1} \times \Z_{p^2}^{\alpha_2} \times \dots \times \Z_{p^k}^{\alpha_k}$, where $\mathbf{v}_i, \mathbf{u}_i\in \Z_{p^i}^{\alpha_i}$, the inner product is defined by
\begin{equation}\label{eq:inner_gen}
\langle \mathbf{v},\mathbf{u} \rangle = p^{k-1} \langle \mathbf{v}_1,\mathbf{u}_1 \rangle +
p^{k-2} \langle \mathbf{v}_2,\mathbf{u}_2 \rangle + \dots +
 \langle \mathbf{v}_k,\mathbf{u}_k \rangle,
\end{equation}
and the Gray-like maps $\varphi$ and $\Phi$ are defined by
\begin{IEEEeqnarray*}{rCl}
 \varphi(\mathbf{v})&=&\big(\mathbf{v}_1|\varphi_2(\mathbf{v}_2)|\ldots |\varphi_k(\mathbf{v}_k)\big) \quad \mbox{and}
 \\
 \Phi(\mathbf{v})&=& \{\mathbf{v}_1\} \times \Phi_2(\mathbf{v}_2) \times  \ldots \times \Phi_k(\mathbf{v}_k).
\end{IEEEeqnarray*}

\end{remark}

\section{Additive $1$-perfect codes}\label{s:perfect}
\begin{definition}
A set $C$ of vertices of a finite metric space is called
an $e$-perfect code if every vertex is at distance $e$ or less from exactly one element of $C$; in other words, if every ball of radius $e$ contains exactly one codeword.
\end{definition}

In this section, we characterize the additive $1$-perfect codes
in the mixed $\Z_p \Z_{p^2}\ldots\Z_{p^k}$ alphabet, $p$ prime, with the distance $d^\diamond$ defined as in Section~\ref{s:prel}.
Once we have the inner product (\ref{eq:inner_gen}), we can define additive codes with the help of check matrices.

\begin{theorem}\label{th:perfect}
  Assume that $A$ is a matrix with rows from $\Z_p^{\alpha_1} \times \Z_{p^2}^{\alpha_2} \times \dots \times \Z_{p^k}^{\alpha_k}$
  such that the first $\gamma_1 \ge 0$ rows are of order $p$,
  the next $\gamma_2 \ge 0$ rows are of order $p^2$, and so on; the last $\gamma_k > 0$ rows are of order $p^k$.
  Assume that all the rows are linearly independent.
  The additive code $\mathcal{C}\subset \Z_p^{\alpha_1} \times \Z_{p^2}^{\alpha_2} \times \ldots \times \Z_{p^k}^{\alpha_k}$ defined by the check matrix $A$ is a $1$-perfect code with respect to the distance $d^\diamond$ if and only if
  for every $i$ from $1$ to $k$,
  \begin{enumerate}
      \item[\rm (i)] $\alpha_i=p^{\gamma_1}p^{2\gamma_2}\ldots p^{(i-1)\gamma_{i-1}}(p^{i\gamma_{i,k}}-p^{(i-1)\gamma_{i,k}})/(p^i-p^{i-1})$, where $\gamma_{i,k}=\gamma_i+\dots+\gamma_k$;
      \item[\rm (ii)] the order of each of $\alpha_i$ columns of $A$ corresponding to $\Z_{p^i}$ is $p^i$;
      \item[\rm (iii)] the $\alpha_i$ columns of the matrix $A$ corresponding to $\Z_{p^i}$ are mutually non-collinear.
  \end{enumerate}
\end{theorem}
\begin{proof}
{\em Only if.} Assume that $\mathcal{C}$ is a $1$-perfect code. It is straightforward from the definition of a $1$-perfect code that a codeword cannot have $\mathrm{wt}^\diamond$ weight $1$ or $2$. Now (ii) is straightforward as
if the $j$th column over $\Z_{p^i}$ has order smaller than $p^i$, then the word with $p^{i-1}$ in the $j$th position and $0$s in the other must be a codeword of weight $1$ (in the case $i=1$) or $2$ (for $i>1$), leading to a contradiction.

(iii) is also clear because if two columns $\mathbf x$ and $\mathbf y$ of maximal order are collinear, then $\mathbf x=\kappa \mathbf  y$ for some $\kappa$ of order $p^i$ in $\Z_{p^i}$, which results in a weight-$2$ codeword with $1$ in the position of $\mathbf x$, $-\kappa$ in the position of $\mathbf y$ and $0$ in the other positions.

Let us prove (i).
The total number of different possible columns of order $p^i$ over $\Z_{p^i}$ is
$p^{\gamma_1}p^{2\gamma_2}\ldots p^{(i-1)\gamma_{i-1}}(p^{i\gamma_{i,k}}-p^{(i-1)\gamma_{i,k}})$
(the first $\gamma_1+ \dots +\gamma_{i-1}$ elements are arbitrary, with the restriction on order; the last $\gamma_i+ \dots +\gamma_{k}$ are arbitrary of order at most $p^i$, but there is at list one element of order exactly $p^i$). To obtain the maximum number of non-collinear columns, we divide this number by the number $p^i-p^{i-1}$ of elements of maximum order in $\Z_{p^i}$ and obtain the formula in (i). So, $\alpha_i$ cannot exceed this value. On the other hand, calculating the cardinality of a radius-$1$ ball $B$ gives
\begin{IEEEeqnarray}{rCl}
|B|&=&
1+\sum_{i=1}^k (p^i-p^{i-1})\alpha_i \nonumber \\
&\le&
1+\sum_{i=1}^k p^{\gamma_1}p^{2\gamma_2}...p^{(i-1)\gamma_{i-1}}(p^{i\gamma_{i,k}}-p^{(i-1)\gamma_{i,k}}) \label{eq:qwe}
\\
&=&
p^{\alpha_1}p^{2\alpha_2}...p^{k\alpha_{k}}.\nonumber
\end{IEEEeqnarray}
From the definition of the $1$-perfect code,
$|B|\cdot|\mathcal{C}|$ coincides with the size of the space;
so, (\ref{eq:qwe}) is satisfied with equality. (i) follows.

{\em If.}
From (ii) and (iii), it is easy to see that for different
$\mathbf e$ of weight $1$ from $\Z_p^{\alpha_1} \times \Z_{p^2}^{\alpha_2} \times \dots \times \Z_{p^k}^{\alpha_k}$, the syndromes $A{\mathbf e}$ are different.
From (iii) and numerical considerations above, we see that $A{\mathbf e}$ exhausts all possible nonzero syndromes from $\Z_p^{\gamma_1} \times \Z_{p^2}^{\gamma_2} \times \ldots \times \Z_{p^k}^{\gamma_k}$. It follows that for every
$\mathbf u$ from $\Z_p^{\alpha_1} \times \Z_{p^2}^{\alpha_2} \times \ldots \times \Z_{p^k}^{\alpha_k}$ there is unique $\mathbf e$ of weight $1$ such that $A{\mathbf e}=A{\mathbf u}$; i.e.,
${\mathbf u}-{\mathbf e}$ is a unique codeword at distance $1$ from ${\mathbf u}$. Hence, the code is $1$-perfect, by the definition.
\end{proof}

\begin{corollary}\label{c:1perfect}
For every integer $k>0$,
$\gamma_1\ge 0$, \ldots,
$\gamma_{k-1}\ge0$, $\gamma_k>0$,
there exists an additive $1$-perfect code in
$\Z_p^{\alpha_1} \times \Z_{p^2}^{\alpha_2} \times \ldots \times \Z_{p^k}^{\alpha_k}$,
where
\begin{IEEEeqnarray*}{r}
 \alpha_i=p^{\gamma_1}p^{2\gamma_2}\ldots p^{(i-1)\gamma_{i-1}}(p^{i\gamma_{i,k}}-p^{(i-1)\gamma_{i,k}})/(p^i-p^{i-1}),\\
\gamma_{i,k}=\gamma_i+\dots+\gamma_k.
\end{IEEEeqnarray*}
Moreover, all $1$-perfect codes in $\Z_p^{\alpha_1} \times \Z_{p^2}^{\alpha_2} \times \dots \times \Z_{p^k}^{\alpha_k}$ are obtained from each other by monomial transformations, i.e., by permutations of coordinates within each of the $k$ groups and multiplication of each coordinate by a unit of the corresponding ring.
\end{corollary}

\begin{example}
Let $p=3$, $k=3$, $\gamma_1=1$,  $\gamma_2=0$,  $\gamma_3=1$. The following matrix is a check matrix of a $1$-perfect code in $\Z_{3}^4\times \Z_{9}^3\times \Z_{27}^3$:
$$ \left(\begin{array}{cccc|ccc|ccc}
   0&1&1&1   &  0&3&6 & 0&9&18
    \\ \hline
   1&0&1&2  &  1&1&1 & 1&1& 1
\end{array}\right).
$$
The size of a radius-$1$ ball is
$1+\alpha_1\cdot(3-1)+\alpha_2\cdot(3^2-3^1)+\alpha_3\cdot(3^3-3^2) = 1+ 4 \cdot 2 + 3 \cdot 6 + 3\cdot 18 =81 =3^{\gamma_1}\cdot 9^{\gamma_2}\cdot 27^{\gamma_3}$.
\end{example}

The following corollary provides an additive analogue
of a one-weight linear code in the Hamming space, known as the simplex code.

\begin{corollary}\label{c:simplex}
Under the assumption and notation
of Corollary~\ref{c:1perfect},
 the dual $\mathcal{C}^\perp$ of an additive $1$-perfect code $\mathcal{C}$ in
 $\Z_p^{\alpha_1} \times \Z_{p^2}^{\alpha_2} \times \ldots \times \Z_{p^k}^{\alpha_k}$
 is an additive one-homogeneous-weight code with the non-zero homogeneous weight $p^{\gamma -1}$,
 where $\gamma=1\gamma_1+2\gamma_2+\ldots +k\gamma_k$.
\end{corollary}

\begin{proof}
Applying the Gray map $\varphi$
to $\mathcal{C}$ we get a code of length
$n=(p^{\gamma}-1)/(p-1)$
size $p^{n-\gamma}$, and minimum distance $3$;
i.e., a code with the same parameters as the Hamming code. From Remark \ref{remark1}, it follows that $\Phi(\mathcal{C}^\perp)$ has the same weight distribution as the dual of the Hamming code, which is a one-weight (simplex) code of non-zero weight $p^{\gamma -1}$ \cite{MWS}.
Since the weight distribution of $\Phi(\mathcal{C}^\perp)$ is the same as the homogeneous weight distribution of $\mathcal{C}^\perp$, the statement follows.
\end{proof}

\begin{remark}
Theorem~\ref{th:perfect} generalizes the result of \cite{BorRif:1999} concerning the classification of the additive $1$-perfect codes in the mixed $\Z_2\Z_4$ alphabet, to the arbitrary $\Z_p\Z_{p^2}\ldots\Z_{p^k}$ case.
However, the classification of the non-equivalent $p$-ary $1$-perfect codes that can be obtained from additive $1$-perfect codes using the Gray map  (which was also completed in \cite{BorRif:1999} for the $\Z_2\Z_4$ case) remains an open problem, interesting even in the partial cases, e.g., $k=2$.
Similar questions are actual for the binary codes with dual parameters that are obtained by the $\varphi$ map from the codes considered in Corollary~\ref{c:simplex}.
Recently, some results in this direction were obtained for
$\Z_{2^k}$-linear Hadamard codes (the codes formally dual to the extended $1$-perfect binary codes)
\cite{FVV:Z2s-kernel}.
\end{remark}

\providecommand\href[2]{#2} \providecommand\url[1]{\href{#1}{#1}}
  \def\DOI#1{{\small {doi}:
  \href{http://dx.doi.org/#1}{#1}}}\def\DOIURL#1#2{{\small{doi}:
  \href{http://dx.doi.org/#2}{#1}}}

\end{document}